\documentclass[twocolumn,twoside]{IEEEtran}

\usepackage{xcolor,soul,framed} 

\colorlet{shadecolor}{yellow}

\usepackage[pdftex]{graphicx}
\graphicspath{{../pdf/}{../jpeg/}}
\DeclareGraphicsExtensions{.pdf,.jpeg,.png}

\usepackage[cmex10]{amsmath}
\usepackage{array}
\usepackage{mdwmath}
\usepackage{mdwtab}
\usepackage{eqparbox}
\usepackage{url}
\usepackage{flushend,cuted}
\usepackage[noend]{algpseudocode}
\usepackage{algorithmicx,algorithm}
\usepackage{subfigure}
\usepackage{caption}
\usepackage{cite}
\usepackage{xcolor}
\usepackage{epstopdf}
\usepackage{cases}
\usepackage{amscd,textcomp,amssymb,epsfig,graphics,epsf,color,amsmath,balance,cite}
\usepackage{multirow}
\usepackage{booktabs}
\usepackage{stfloats}
\usepackage{setspace}
\usepackage{graphicx}
\usepackage{amsmath, amsthm, amsfonts, amssymb, amsbsy}
\usepackage{dsfont}
\usepackage{color}
\usepackage{bm}
\usepackage{tabu}                     
\usepackage{multirow}                 
\usepackage{multicol}                 
\usepackage{multirow}                
\usepackage{float}                    
\usepackage{makecell}                 
\usepackage{booktabs}                 

\begin{document}

\title{\LARGE{CSI-Free Optimization of Reconfigurable Intelligent Surfaces with Interference by Using Multiport Network Theory}}

\author{ \IEEEauthorblockN{Andrea Abrardo\IEEEauthorrefmark{1}} \\
\IEEEauthorblockA {\IEEEauthorrefmark{1} University of Siena, Siena, Italy \\
Email: abrardo@unisi.it}\\
}

\author{Andrea~Abrardo \IEEEmembership{Senior~Member,~IEEE}
\thanks{A. Abrardo is with the University of Siena and CNIT, Italy (e-mail: abrardo@unisi.it). The work is supported by the 6G SHINE European project. I would like to express my sincere gratitude to Professor Marco Di Renzo and Professor Alberto Toccafondi for their invaluable contributions to this work. The collaboration with Marco Di Renzo in several past and present research activities in the field of RIS modeling and optimization has been instrumental in shaping the direction of this research. The support of Alberto Toccafondi allowed to enrich my understanding and expertise in electromagnetic modeling.
}
}

\maketitle

\begin{abstract}
Reconfigurable Intelligent Surfaces (RIS) will play a pivotal role in next-generation wireless systems. Despite efforts to minimize pilot overhead associated with channel estimation, the necessity of configuring the RIS multiple times before obtaining reliable Channel State Information (CSI) may significantly diminish their benefits. Therefore, we propose a CSI-free approach that explores the feasibility of optimizing the RIS for the uplink of a communication system in the presence of interfering users without relying on CSI estimation but leveraging solely some a priori statistical knowledge of the channel. In this context, we consider a multiport network model that accounts for several aspects overlooked by traditional RIS models used in Communication Theory, such as mutual coupling among scattering elements and the presence of structural scattering. The proposed approach targets the maximization of the average achievable rate and is shown to achieve performance that, in some cases, can be very close to the case where the RIS is optimized leveraging perfect CSI.
\end{abstract}


\vspace{-0.2cm}
\section{Introduction}\label{Intro}

Reconfigurable Intelligent Surfaces (RIS) are considered a pivotal technology for enabling the concept of the \emph{smart radio environment} in the context of next-generation wireless systems, particularly those operating at mmWave frequencies or even higher \cite{LiaNieTsiPitIoaAky:18,DiRenzo:19b,WuZhaZheYouZha:20,DiRenzo2020_JSAC}. In a smart radio environment, the wireless system goes beyond conventional communication device parameters, with the environment itself (e.g., the channel) becoming a variable that can be optimized \cite{Yuanwei2020,DiRDanXiDeRTre:2020,bjornson2022reconfigurable}. In RIS-aided communications, precise estimation of the cascaded channel (from the UE to RIS and from the RIS to the BS) is essential for effective phase-shift design and harnessing the benefits offered by RIS technology. However, this task is complex, mainly attributed to the passive characteristics of the RIS and the complexities posed by high-dimensional channels~\cite{9328501, Demir2022is, Demir2022Exploiting}. Indeed, since all RIS elements are passive, i.e., they cannot transmit, receive, or process any pilot signal, CSI estimation can only be performed by a central controller (e.g., the BS) at the expenses of a huge overhead needed to estimate the channel with respect to standard MIMO systems. Accordingly, CSI estimation has been the subject of several studies in recent years, with the goal of minimizing the number of pilot RIS configurations required for accurate estimation \cite{9328485, Gui2022Channel, 9053695,Chen2023Channel}. The presence of electromagnetic interference, which cannot be ignored, particularly with very large RIS surfaces, can further complicate the CSI problem if appropriate countermeasures are not implemented, as shown in ~\cite{Electromagnetic2022de, Wen-Xuan-CommLett_2023}.

Despite efforts to minimize pilot overhead associated with channel estimation, the need to configure the RIS multiple times before obtaining reliable CSI may significantly diminish their benefits, especially if the number of individually reconfigurable elements is excessively large or if an inefficient channel estimation algorithm is utilized \cite{ABR, Zappone2020}. Furthermore, considering the dynamic nature of the wireless channel and user mobility, reconfiguring a RIS within a timescale aligned with the coherence time of the environment (e.g., the channel) can be challenging, particularly in highly dynamic environments \cite{WuZhaZheYouZha:20}.

To address these open research issues, some authors have recently delved into research on RIS-aided systems that do not strictly depend on perfect CSI knowledge, e.g., see  \cite{ABR,WuZhang2020, Abrardo_two, Chaaban2020, Cunhua2020, DarMas:20}. Specifically, in \cite{WuZhang2020}, the phase shifts of the RIS and the transmit beamforming/precoding vector at the access point are optimized using a two-timescale beamforming optimization algorithm. The RIS is optimized based on the statistical CSI of all links, which is operated on a long time-scale, effectively reducing the CSI overhead. Simultaneously, the transmit beamforming/precoding vectors at the access point are designed based on the instantaneous CSI of the effective channels, given the optimized RIS configuration. Following a different direction, in \cite{ABR,Abrardo_two}, the RIS design is conducted offline assuming \textit{a priori} statistical knowledge of UE locations. This involves a two-phase optimization process, encompassing an offline phase (long-term and sporadic) and an online phase (short-term and more frequent). During the online phase, communication between the BS and the UEs is optimized without interaction with the RISs. The key advantage of this approach is that information necessary for RIS optimization can either be known beforehand (e.g., knowledge of UEs confined within a specific area) or can be occasionally learned during system operation, leveraging a localization infrastructure.

In all the previously mentioned works, and in the majority of works addressing RIS optimization for communications and/or CSI estimation, a RIS is characterized as a planar array comprising a given number of reflective elements strategically positioned at sub-wavelength intervals. The impedance of each element can be finely adjusted to introduce a controllable phase-shift to the incident wave before reflecting it. By optimizing the phase-shift pattern across the RIS, it becomes possible to manipulate the reflected wavefront, directing it into a beam aimed at the intended receiver. These models are not always electromagnetically consistent, since they do not consider several aspects that play an important role in characterizing the operation of a realistic RIS \cite{RenzoDT22, direnzo2022digital, AbeywickramaZWY20}. Recent results highlight, in fact, the critical need of using realistic reradiation models \cite{Galdi2023}, which result in a strong interplay between the surface-level optimization of RISs and the element-level design of the RIS elements \cite{RafiqueHZNMRDY23}. In this context, multiport network theory has been proved to be a suitable approach for modeling and optimizing RIS-aided channels \cite{IvrlacN10}. The motivation for using multiport network theory lies in its inherent capability of modeling the electromagnetic mutual coupling between closely-spaced radiating elements, which is a key distinguishing feature of RISs, especially for realizing advanced wave transformations at a high power efficiency \cite{RenzoZDAYRT20,10213362, li2023tunable}.

The first multiport model for RIS-aided channels was introduced in \cite{DR1}, assuming a minimum scattering radiating element configuration. The model employed thin wire dipoles as scattering elements for analytical tractability, aligning with the discrete dipole approximation \cite{math10173049}. A similar model based on the coupled dipoles formalism is presented in \cite{FaqiriSASIH23}. The distinctive feature of the communication model in \cite{DR1} is its non-linearity, arising from the tunable impedances, leading to a system response affected by electromagnetic mutual coupling between closely spaced RIS elements. Subsequently, this multiport model has been applied in various works considering different communication scenarios, such as Single-Input Single-Output (SISO), Multiple-Input Single-Output (MISO), and Multiple-Input Multiple-Output (MIMO), and different channel models \cite{DR2, ABR, DR3, abs-2305-12735, akrout2023physically, abs-2306-03761, zheng2023impact,9514409}.

In two recent works \cite{abrardo_E}, \cite{abrardo_S}, an easily applicable multiport network model is proposed, based on the S and Z parameters representation of the network. This model reveals the approximations inherently considered in classical RIS models used in communication theory that assume the RIS as an ideal scatterer. These classical models overlook the effects of: (\emph{i}) electromagnetic mutual coupling between scattering elements; (\emph{ii}) correlation between the phase and amplitude of the reflection coefficient; (\emph{iii}) the presence of a structural scattering component, leading to an unwanted specular component of the reflected wavefront.

In all these works dealing with multiport network models, the channel is assumed to be perfectly known at the central controller, and RIS optimization is carried out accordingly.

\subsection{Contribution}

In this work, we adopt the CSI-free approach, exploring the feasibility of optimizing the RIS without relying on CSI estimation. Similar to other works proposing methods for cascade channel estimation, such as \cite{Demir2022is}, \cite{Wen-Xuan-CommLett_2023}, and \cite{xuan2023}, we presume a priori statistical knowledge of the channel, specifically the correlation matrix. The two main novel contributions of this work are listed below.

\begin{itemize}
\item
 Instead of utilizing the knowledge of the channel correlation matrix to develop an efficient channel estimation, we employ it directly for RIS optimization, as proposed in \cite{WuZhang2020}, \cite{ABR}, and \cite{Abrardo_two}. However, unlike these works, we do not rely on a Monte Carlo approach for RIS optimization, which can be computationally very intensive. Instead, we employ an approach that optimizes the RIS directly based solely on the correlation matrix of the channel. This approach is generalized to accommodate the presence of a certain number of interferers in the form of other transmitting nodes.
\item
The proposed approach fot CSI-free RIS optimization leverages the multiport network model proposed in \cite{abrardo_S}. As discussed earlier, this model allows for a more realistic characterization of the RIS compared to the ideal models considered so far in the field of CSI estimation or CSI-free RIS optimization. 
\end{itemize}
The offline CSI-free RIS optimization proposed in this paper is carried out assuming that in the channel between the Base Station (BS) and the RIS, the presence of natural scattering can be neglected and that the BS and the RIS are positioned in fixed and known locations. The first assumption arises from the consideration that the BS and the RIS can, in many cases, be optimally deployed to have a very strong Line-of-Sight (LOS) component. However, in the model considered for simulations, we will take into account the presence of multipath also in the BS-RIS link. In the Results section, we will demonstrate the validity of the proposed approach and how the CSI-free approach allows for achieving good performance in terms of achievable throughput in an uplink communication scenario. It is also demonstrated how the use of simplified models that assume the Reconfigurable Intelligent Surface (RIS) as an ideal scatterer results in a significant performance loss due to model mismatch.

\subsection{Paper Outline and Notation}
The rest of this paper is organized as follows. In Section~\ref{sec:system_model}, we introduce the model adopted for the system, including the propagation models and the communication scenario. In Section III, the multiport network model for the Reconfigurable Intelligent Surface (RIS) characterization is proposed. In Section IV, we provide an iterative algorithm with provable convergence that allows carrying out the CSI-free RIS optimization. Numerical results are presented in Section V, while conclusions are drawn in Section VI.

{\sl Notation}: Unless otherwise specified, matrices are denoted by bold uppercase letters (i.e., $\mathbf{X}$), vectors are represented by bold lowercase letters (i.e., $\mathbf{x}$), and scalars are denoted by normal font (i.e., $x$). $(\cdot)^{\mathrm{T}}$, $(\cdot)^{\mathrm{H}}$ and $(\cdot)^{-1}$ stand for the transpose, Hermitian transpose and inverse of the matrices. The symbol $\odot$ represents the Hadamard (element-wise) product while with $\text{diag}\left(\mathbf{x}\right)$ we mean the diagonal matrix obtained from the element of vector $\mathbf{x}$. The notation $||\mathbf{x}||$ signifies the Euclidean norm of the vector $\mathbf{x}$, and $\mathbb{E}\{\cdot\}$ represents the statistical expectation.

\vspace{-0.1cm}
\section{System Model}\label{sec:system_model}

We consider a wireless scenario with $N_u$ single-antenna UEs that must establish a communication link with an $N$-element BS. The BS antennas are arranged in a uniform planar array (UPA) with $N_H$ rows and $N_V$ columns, resulting in $N = N_HN_V$. In this environment, a Reconfigurable Intelligent Surface (RIS) is present with the goal of supporting the communication of one of the UEs, referred to as the reference user. The RIS is equipped with $M$ passive reconfigurable elements, forming a UPA with $M_H$ rows and $M_V$ columns, where $M=M_HM_V$. Without loss of generality, we assume that the first UE serves as the reference user. Hence, in the considered scenario, the objective of the RIS is to maximize the rate of the first UE, while users $i$, where $i = 2,\ldots,N_u$, act as interferers. 

\begin{figure}[t]
\setlength{\abovecaptionskip}{-0.1cm}
\setlength{\belowcaptionskip}{-0.3cm}   \begin{center}
\includegraphics[scale=0.28]{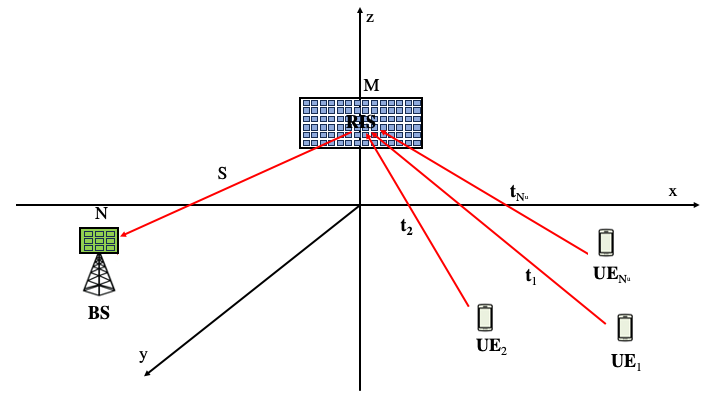}
\end{center}
\captionsetup{font=small}
\caption{RIS-aided communication system.}
\label{Fig1}
\end{figure}

Since we are interested in finding the optimal Reconfigurable Intelligent Surface (RIS) configuration, we neglect in what follows the direct links between the User Equipments (UEs) and the Base Station (BS) that are not under the control of the RIS. On the other hand, as discussed in most previous works on RIS, e.g., see \cite{Perovic, Chien21, ABR}, in the presence of a strong BS-UE direct link, the impact and contribution of the RISs are usually not very significant. Hence, optimizing the RISs to take into account this possibility typically has no significant benefit.

Accordingly, denoting by $\mathbf{t}_i \in \mathbb{C}^{M \times 1}$ the channel between UE $i$ and the RIS, and by $\mathbf{S} \in \mathbb{C}^{N \times M}$ the channel between the RIS and the BS, we can express the received vector $\mathbf{y} \in \mathbb{C}^{N \times 1}$ as follows:
\begin{align}\label{eq1}
& \mathbf{y} = \mathbf{S} \boldsymbol{\Delta}(\mathbf{z}) \sum\limits_{i = 1}^{N_u}\mathbf{t}_i s_i + \mathbf{n}.
\end{align}
Here, $\boldsymbol{\Delta}(\mathbf{z}) \in \mathbb{C}^{M \times M}$ is the RIS reflecting matrix, dependent on a vector $\mathbf{z} \in \mathbb{C}^{M \times 1}$ of tunable parameters, $s_i$ represents the transmitted symbol by UE $i$, and $\mathbf{n} \in \mathbb{C}^{N \times 1}$ is the thermal noise with variance $\sigma^2_n$.

Concerning the matrix $\mathbf{S}$, it is assumed to be perfectly known from the network deployment. On the other hand, the Base Station (BS) and the Reconfigurable Intelligent Surface (RIS) are typically in fixed positions and arranged to establish a strong LOS that depends solely on their respective positions. Therefore, the channel can be assumed to be deterministic and known. In the model considered for simulations, we will take into account the presence of a certain amount of multipath in the BS-RIS link, showing that this presence does not impact the validity of this assumption, provided, of course, that the LOS component overwhelms the scattering component.

Conversely, the channels $\mathbf{t}_i$ are not known to the BS, and for this reason, the optimization of the Reconfigurable Intelligent Surface (RIS) requires either a direct estimation or prior knowledge of certain statistical parameters characterizing these channels. This latter approach is considered in this paper as it avoids the heavy overhead of channel estimation, as discussed in Section \ref{Intro}. In particular, we assume to know the correlation matrices:
\begin{align}\label{eq2}
& \mathbf{R}_{\mathbf{t}_i} = \mathbb{E}\left(\mathbf{t}_i \mathbf{t}^H_i\right)
\end{align}
with $i = 1,\dots,N_u$ and we denote by $\mathbf{x} = \mathbf{t}_1 s_1$, the intended signal, and by $\mathbf{w}_i = \mathbf{t}_i s_i$, with $i > 1$ the interfering users. Then we introduce:
\begin{align}\label{eq3}
& \mathbf{R}_{\mathbf{x}} =  \mathbb{E}\left(\mathbf{x} \mathbf{x}^H\right) =  \sigma_1^2 \mathbf{R}_{\mathbf{t}_1} \nonumber \\
& \mathbf{R}_{\mathbf{w}_i} =  \mathbb{E}\left(\mathbf{w}_i \mathbf{w}_i^H\right) =  \sigma_i^2 \mathbf{R}_{\mathbf{t}_i} \quad i = 2,\ldots,N_u
\end{align}
where $\sigma_i^2 =   \mathbb{E}\left(|s_i|^2\right)$ is the transmitted power of the $i$-th UE.  It is worth noting that this assumption is often employed to support Channel State Information (CSI) estimation in previous works, such as \cite{Abrardo_two, Sangui1, Wen1}. In our paper, we utilize this a-priori information to optimize the Reconfigurable Intelligent Surface (RIS) without involving any channel estimation phase. \\

\section{RIS Models}\label{Models}

\subsection{Communication Theory (CT) Model}

The model traditionally used to characterize RIS and integrate them into a communication system is to represent them as surfaces composed of ideal scatterers capable of imparting a phase shift on the impinging signal. This model, which we will refer to as the Communication Theory model (CT) hereafter, can be represented by a diagonal reflection matrix with elements:
\begin{equation}\label{eq:TFZ_CT2}
\left[\boldsymbol{\Delta}_{CT}(\mathbf{z})\right]_{m,m} = K_0 \Gamma_m
\end{equation}
where $K_0$ is a constant depending on the geometry of the scatterer, and $\Gamma_m = e^{j\phi_m}$ is the load reflection coefficient which can be changed by acting on the tunable parameter $z_m$.

\subsection{Multiport Network (MP) Model}
Referring to the multiport $Z$-parameters representation of an end-to-end channel matrix involving an RIS presented in \cite{abrardo_E}, we have:
\begin{equation}\label{eq:TFZ}
\boldsymbol{\Delta}_{MP}(\mathbf{z}) = -2Y_0 (\mathbf{Z}_{SS}+r_0 \mathbf{I}_M+\mathbf{Z}_R)^{-1}.
\end{equation}
Here, $\mathbf{Z}_{SS} \in \mathbb{C}^{M \times M}$ represents the matrix of self and mutual impedances of the RIS, $\mathbf{Z}_R = \text{diag}(\mathbf{z})$ where $\mathbf{z}$ are tunable impedances connected at the RIS ports, $r_0$ is a small parasitic resistance, $\mathbf{I}_M$ is the identity matrix, and $Y_0$ is a reference admittance which depends, among other factors, on the reference impedance $Z_0$. For simplicity, it is assumed that $Z_0$ is the same for all ports and equal to $Z_0 = 50\Omega$. Note that \eqref{eq:TFZ} has dimensions of $\Omega^{-2}$, which is consistent with the fact that when using the Z-parameter representation of the multiport, $\mathbf{S}$ and $\mathbf{t}_i$ in \eqref{eq1} are impedance matrices with dimensions of $\Omega$.

\subsection{Comparison between CT and MP models}

The CT model in \eqref{eq:TFZ_CT2} can be seen as a particular case of the more general MP model in \eqref{eq:TFZ}, as broadly discussed in \cite{abrardo_E}. It is indeed straightforward to derive the CT model from the MP one my making the following assumptions. First we neglect the parasitic resistance $r_0$ and impose $\mathbf{Z}_{SS} = Z_0 \mathbf{I}_M$. In this manner, beyond ignoring the mutual coupling, the scattering matrix of the RIS becomes zero, and the effect of the RIS is solely to introduce a pure phase shift to the incident signal. This simplified model is denoted by ideal Multi-Port (iMP) which, from \eqref{eq:TFZ}, yields a diagonal RIS reflection matrix with elements:
\begin{equation}\label{eq:TFZ_CT1}
\left[\boldsymbol{\Delta}_{iMP}(\mathbf{z})\right]_{m,m} = -2Y_0 ({Z}_{0}+z_m)^{-1}. 
\end{equation}
Hence, it is easy to verify that:
\begin{equation}\label{eq:TFZ_CT1bis}
 \boldsymbol{\Delta}_{iMP}(\mathbf{z}) = \boldsymbol{\Delta}_{CT} (\mathbf{z}) - \frac{Y_0}{{Z}_{0}} \mathbf{I}_M,
\end{equation}
when $K_0 = \frac{Y_0}{Z_0}$ and $\Gamma_m({z}_m) = \frac{jb_m-Z_0}{jb_m+Z_0}$. The above equation allows to shed light on a third important approximation which is inherent in the CT model, in which $\Delta_{\mathbf{S}} = - \frac{Y_0}{{Z}_{0}} \mathbf{I}_M$ is not considered. This term corresponds to the scattered signal obtained when the the RIS is terminated with impedances $z_m = Z_0$, i.e., when the load reflection coefficient is zero, and hence the CT model would entails zero scattered energy. This structural component causes the RIS to behave like a scatterer that does not introduce any phase shift on the incident signal, thereby introducing a specular component that, as we will see, cannot be ignored without significantly deteriorating performance in certain situations.


\section{RIS Optimization}\label{RIS_opt}

\subsection{Problem Formulation}

From now on, we will consider the case where the RIS is terminated with a purely reactive impedance, which is the case of interest where we have maximum signal reflection. Therefore, we consider $\mathbf{z} = j\mathbf{b}$. For simplicity, we denote $\boldsymbol{\Delta}(\mathbf{b}) = \boldsymbol{\Delta}(j\mathbf{b})$. Hence, denoting by $\boldsymbol{\Phi} = \mathbf{S} \boldsymbol{\Delta}(\mathbf{b})$ and emphasizing the effect of the intended UE with respect to the others, we can rewrite \eqref{eq1} as follows:
\begin{align}\label{eq4}
& \mathbf{y} = \boldsymbol{\Phi} \mathbf{x} + \boldsymbol{\Phi} \sum\limits_{i = 2}^{N_u}\mathbf{z}_i  + \mathbf{n}.
\end{align}
The BS estimates the transmitted signal $s_1$ by using the combining vector $\mathbf{v} \in \mathbb{C}^{N \times 1}$ to obtain:
\begin{align}\label{eq5}
& \hat{s}_1 = \mathbf{v}^H\mathbf{y} = \mathbf{v}^H\boldsymbol{\Phi} \mathbf{x} + \mathbf{v}^H\boldsymbol{\Phi} \sum\limits_{i = 2}^{N_u}\mathbf{z}_i  +\mathbf{v}^H \mathbf{n}.
\end{align}

Considering the worst-case bound for the evaluation of the rate based on the assumption of uncorrelated additive noise given in \cite{Hassibi2003}, we can write the following expression for the ergodic achievable uplink rate: 
\begin{align}\label{ERG_Rate}
R & = \mathbb{E}_\mathbf{x}\left[\text{log}_2\left(1+\gamma(\mathbf{x},\boldsymbol{\Phi})\right)\right] 
\end{align}
where $\mathbb{E}_\mathbf{x}$ is the expectation with respect to $\mathbf{x}$ and 
\begin{align}\label{ERG_Gamma}
\gamma(\mathbf{x},\boldsymbol{\Phi}) & = \frac{\mathbf{v}^H\boldsymbol{\Phi} \mathbf{x} \mathbf{x}^H \boldsymbol{\Phi}^H \mathbf{v}}{\mathbb{E}\left(\sum\limits_{i=2}^{N_u}\mathbf{v}^H \boldsymbol{\Phi} \mathbf{z}_i \mathbf{z}_i^H \boldsymbol{\Phi}^H \mathbf{v} + \mathbf{v}^H\mathbf{v}\sigma^2_n\right)}\\
& = \frac{\mathbf{v}^H \boldsymbol{\Phi} \mathbf{x} \mathbf{x}^H \boldsymbol{\Phi}^H \mathbf{v}}{\mathbf{v}^H \left(\boldsymbol{\Phi} \sum\limits_{i=2}^{N_u} R_{\mathbf{w}_i} \boldsymbol{\Phi}^H + \sigma^2_n \mathbf{I}_N\right)\mathbf{v}} \nonumber
\end{align}
In order to provide a CSI-free RIS design, the RIS could therefore be designed with the aim of maximizing:
\begin{align}\label{ERG_integral}
\int\limits_{\mathbf w}\gamma(\mathbf{w},\boldsymbol{\Phi}) f_\mathbf{x}(\mathbf{w}) d \mathbf{w}
\end{align}
Where $f_\mathbf{x}(\mathbf{w})$ is the probability density function (pdf) of $\mathbf{x}$, i.e., aimed at maximizing the average rate. This approach has already been proposed in previous works, for example, see \cite{ZhaWuZhaZha:19} and \cite{ABR}, for traditional RIS models. However, it has the drawback of requiring knowledge of the full channel statistics. Moreover, since the integral in \eqref{ERG_integral} cannot be evaluated in closed form, it is necessary to rely on a Monte Carlo approach. To overcome these challenges, we consider the upper bound $R_b \geq R$ derived from the concavity of the logarithm function, where:
\begin{align}\label{ERG_Rate1}
& R_b = \text{log}_2\left\{1+\mathbb{E}_\mathbf{x}\left[\gamma\left(\mathbf{x},\boldsymbol{\Phi}\right)\right]\right\}.
\end{align}
From \eqref{ERG_Gamma} we have:
\begin{align}\label{ERG_SINR}
\mathbb{E}_\mathbf{x}\left[\gamma\left(\mathbf{x},\boldsymbol{\Phi}\right)\right] & = \frac{\mathbf{v}^H \boldsymbol{\Phi} \mathbf{R}_{\mathbf{x}} \boldsymbol{\Phi}^H \mathbf{v}}{\mathbf{v}^H \left(\boldsymbol{\Phi} \sum\limits_{i=2}^{N_u} R_{\mathbf{w}_i} \boldsymbol{\Phi}^H + \sigma^2_n \mathbf{I}_N\right)\mathbf{v}}  
\end{align}
for which it is only necessary to know the channel correlation.

To further elaborate, we are then in the position to formulate the optimal CSI-free RIS design as:
\begin{align}\label{Problem1}
\mathbf{b}^* & = \text{arg}\max \limits_{\mathbf{b}} \frac{\mathbf{v}^H \boldsymbol{\Phi}(\mathbf{b}) \mathbf{R}_{\mathbf{x}} \boldsymbol{\Phi}(\mathbf{b})^H \mathbf{v}}{\mathbf{v}^H \boldsymbol{\Phi}(\mathbf{b}) \sum\limits_{i=2}^{N_u} R_{\mathbf{w}_i} \boldsymbol{\Phi}(\mathbf{b})^H \mathbf{v} + \mathbf{v}^H\mathbf{v}\sigma^2_n} \\
&\quad  \textrm{s.t.}
\  \mathbf{\Phi}\in\mathcal{F} \nonumber
\end{align}
where the feasible set 
\begin{equation}
\mathcal{F}=\{\mathbf{\Phi} \in \mathbb{C}^{1 \times M}\vert \mathbf{\Phi} = \mathbf{S} \boldsymbol{\Delta}(\mathbf{b})\}
\end{equation}
enforces the multiport network model, where $\mathbf{b}$ represents the tunable reactances of the RIS, i.e., $\mathbf{b} \in \mathbb{R}^M$. To further elaborate, we first observe that problem \eqref{Problem1} is not convex in $\mathbf{b}$. Hence, we provide in the following an alternative formulation of the same problem which allows to get a local optimum.\\ 
First, to reduce the dimension of the problem we consider the eigen-decomposition $\mathbf{R}_{\mathbf{x}} = \mathbf{U} \mathbf{D} \mathbf{U}^H$ and denote by $r$ the rank of $\mathbf{R}_{\mathbf{x}}$. Then, we denote by $\mathbf{r} \in \mathbb{C}^{r \times 1}$ a vector of i.i.d. Gaussian random variables with $\mathbf{r} \sim \mathcal{N}(\mathbf{0},\mathbf{I}_r)$. Observing that $\mathbf{U} \mathbf{D}^{\frac{1}{2}}\mathbf{r}$ and $\mathbf{x}$ have the same correlation matrix, we rewrite \eqref{eq4} as:
\begin{align}\label{eq4bis}
& \mathbf{y} = \boldsymbol{\Phi} \mathbf{U} \mathbf{D}^{\frac{1}{2}}\mathbf{r} + \boldsymbol{\Phi} \sum\limits_{i = 2}^{N_u}\mathbf{z}_i + \mathbf{n}.
\end{align}
The linear minimum mean square error (LMMSE) estimator of $\mathbf{{r}}$ from $\mathbf{v}^H \mathbf{y}$ is the linear filter $\mathbf{\Lambda}\in\mathbb{C}^{r \times 1}$ designed to minimize the mean square error (MSE) between the vector $\mathbf{r}$ and its MMSE estimate $\mathbf{\hat{r}}= \mathbf{\Lambda}\mathbf{v}^H\mathbf{y}$, i.e.,
\begin{align}\label{r_MMSE}
\mathcal{E}_\mathbf{r} = \mathbb{E}\left\{\left\|\mathbf{\Lambda}\mathbf{v}^H\mathbf{y}-\mathbf{r}\right\|^2\right\}.
\end{align}
The linear estimator that minimizes \eqref{r_MMSE} is then given by:
\begin{align}\label{Lamda_Phi}
\mathbf{\Lambda}(\mathbf{\Phi})\! & = \frac{\!\mathbf{D}^{\frac{1}{2}} \mathbf{U}^H \mathbf{\Phi}^{H} \mathbf{v}}{\mathbf{v}^H \left(\mathbf{\Phi}\mathbf{R}_\mathbf{x}\mathbf{\Phi}^{H}+\boldsymbol{\Phi} \sum\limits_{i=2}^{N_u} R_{\mathbf{w}_i} \boldsymbol{\Phi}^H + \sigma^2_n \mathbf{I}_N\right) \mathbf{v}} , 
\end{align}
where $\mathbf{\Lambda}(\mathbf{\Phi}) \in \mathbb{C}^{r \times 1}$ depends on the RIS phase-shift matrix $\mathbf{\Phi}$. Substituting \eqref{Lamda_Phi} into \eqref{r_MMSE} yields:
\begin{equation}\label{MSE_MMSE}
\mathcal{E}_\mathbf{r}(\mathbf{\Phi}) = \text{tr}\left(\mathbf{I}_r-\mathbf{\Lambda}(\mathbf{\Phi})\mathbf{v}^H\mathbf{\Phi}\mathbf{U}\mathbf{D}^{\frac{1}{2}}\right).
\end{equation}

Denoting $\mathbf{s} = \mathbf{D}^{\frac{1}{2}} \mathbf{U}^H \mathbf{\Phi}^{H} \mathbf{v}$, from \eqref{Lamda_Phi}, \eqref{MSE_MMSE} can be written as:
\begin{align}\label{MSE_MMSE2}
\mathcal{E}_\mathbf{r}(\mathbf{\Phi}) & =\text{tr}\left(\mathbf{I}_r- \frac{\!\mathbf{s}\mathbf{s}^H}{\mathbf{v}^H \left[\mathbf{\Phi}\left(\mathbf{R}_\mathbf{x}+ \sum\limits_{i=2}^{N_u} R_{\mathbf{w}_i}\right) \boldsymbol{\Phi}^H + \sigma^2_n \mathbf{I}_N \right]\mathbf{v}}\right) \nonumber \\
& = r-\frac{\!\mathbf{s}^H\mathbf{s}}{\mathbf{v}^H \left[\mathbf{\Phi}\left(\mathbf{R}_\mathbf{x}+ \sum\limits_{i=2}^{N_u} R_{\mathbf{w}_i}\right) \boldsymbol{\Phi}^H + \sigma^2_n \mathbf{I}_N \right]\mathbf{v}} \nonumber \\
& = r-1 + \frac{1}{\mathbb{E}_\mathbf{x}\left[\gamma\left(\mathbf{x},\boldsymbol{\Phi}\right)\right]+1}
\end{align}
Thus, problem \eqref{Problem1} is equivalent to:
\begin{align}\label{Problem2}
\mathbf{b}^* = & \text{ arg}\min \limits_{\mathbf{b}} \mathcal{E}_\mathbf{r}(\mathbf{\Phi}(\mathbf{b}))   \\
&\quad  \textrm{s.t.}
\  \mathbf{\Phi}\in\mathcal{F} \nonumber 
\end{align}
In other words, the optimal design of the Reconfigurable Intelligent Surface (RIS), aimed at maximizing the ergodic rate for a given combining vector $\mathbf{v}$ without channel estimation, is equivalent to the optimal RIS design for minimizing the estimation error of the channel based on a single observation of the received signal obtained using the same combining vector $\mathbf{v}$. This equivalence holds particularly when channel estimation employs a Linear Minimum Mean Squared Error (LMMSE) estimator.

Problem \eqref{Problem2} is still not convex in $\mathbf{b}$. To address the solution of \eqref{Problem2}, we make a simplifying assumption, neglecting the fact that $\mathbf{\Lambda}(\mathbf{\Phi})$ depends on the value of $\mathbf{\Phi}$. Hence, we express the objective function of \eqref{Problem2} as
\begin{equation} \label{AO1_1}
\mathcal{E}_\mathbf{r}(\mathbf{\Phi}) =\mathcal{E}_\mathbf{r}(\mathbf{\Lambda},\mathbf{\Phi}(\mathbf{b})),
\end{equation}
as if $\mathbf{\Lambda}$ and $\mathbf{\Phi}$ were independent variables.
Under this hypothesis, \eqref{Problem2} can be solved by following an \emph{alternating optimization} (AO) approach. The AO is an iterative algorithm whose  key advantage is that it simplifies the optimization process by breaking it into smaller subproblems, which are easier to solve. The AO approach is especially helpful when the original problem involves complicated interactions or dependencies among the variables.
In particular, we propose a two-step \emph{iterative} algorithm, where $\mathcal{E}_\mathbf{r}$ is alternatively optimized with respect to $\mathbf{\Lambda}$ and $\mathbf{\Phi}$. Being $\mathbf{\Lambda}^{(k)}$ and $\mathbf{\Phi}^{(k)}$ the values found at iteration $k$, at iteration $k+1$ we proceed as follows: 

\begin{enumerate}
\item Having fixed the value of  $\mathbf{\Phi}=\mathbf{\Phi}^{(k)}$, we minimize
  \eqref{AO1_1} by computing  $\mathbf{\Lambda}^{(k+1)}$ as  $\mathbf{\Lambda}(\mathbf{\Phi})$ in \eqref{Lamda_Phi}. This optimization is unconstrained and  is a straightforward application of MSE minimization;  
\item Fixing the value of $\mathbf{\Lambda}$ as $\mathbf{\Lambda}^{(k+1)}$, the MSE takes the expression given in \eqref{MSE_k} and $\mathcal{E}_\mathbf{r}(\mathbf{\Lambda},\mathbf{\Phi})$ is now a convex function of $\mathbf{\Phi}$. The RIS phase-shift matrix  is computed as  the solution of the minimization
\begin{figure*}[b]
\setlength{\abovecaptionskip}{-0.5cm}
\setlength{\belowcaptionskip}{-0.5cm}
\hrulefill
\begin{align}\label{MSE_k}
\mathcal{E}_\mathbf{r}\left(\mathbf{\Lambda}^{(k+1)},\mathbf{\Phi}(\mathbf{b})\right)\!=\!\text{tr}\left\{\mathbf{\Lambda}^{(k+1)}\left[\mathbf{v}^H \left(\mathbf{\Phi}(\mathbf{b})\left(\mathbf{R}_\mathbf{x} +\sum\limits_{i=2}^{N_u} R_{\mathbf{w}_i}\right) \boldsymbol{\Phi}^H(\mathbf{b}) + \sigma^2_n \mathbf{I}_N\right) \mathbf{v}\right]{\mathbf{\Lambda}^{(k+1)}}^{\mathrm{H}}
\!-\!2\Re\left[\mathbf{\Lambda}^{(k+1)}\mathbf{v}^H\mathbf{\Phi}(\mathbf{b})\mathbf{U}\mathbf{D}^{\frac{1}{2}}\right]\!+\!\mathbf{I}_{r}\right\}.
\end{align}
\end{figure*}
\begin{equation}
\begin{aligned} \label{opt_2_1}
&\mathbf{\Phi}^{(k+1)} = \arg \min_{\mathbf{\Phi}}\ \mathcal{E}_\mathbf{r}\left(\mathbf{\Lambda}^{(k+1)},\mathbf{\Phi}\right),\\
&\qquad \qquad \textrm{s.t.}
\  \mathbf{\Phi}\in\mathcal{F}.
\end{aligned}
\end{equation}

\end{enumerate}

Regarding the convergence of the proposed technique we can observe that, since in both steps we minimize the MSE, at each iteration the MSE either decreases or reaches a point where it remains unchanged. Given that the MSE is a positive value, the procedure will ultimately converge to a local optimum. 

\subsection{Solution of \eqref{opt_2_1}}
\subsubsection{Solution for the MP case}
The main difficulty posed by \eqref{opt_2_1} is the nonlinearity of $\mathbf{\Phi}(\mathbf{b}) = \mathbf{S}\Delta(\mathbf{b})$, where $\Delta(\mathbf{b})$ is defined in \eqref{eq:TFZ}. To elaborate, for the sake of notation convenience, let us introduce $Q$ and $\mathbf{q} \in \mathbb{C}^{r \times 1}$ as:
\begin{align}\label{notations}
Q & = \mathbf{v}^H \left(\mathbf{\Phi}(\mathbf{b})\left(\mathbf{R}_\mathbf{x} +\sum\limits_{i=2}^{N_u} R_{\mathbf{w}_i}\right) \boldsymbol{\Phi}^H(\mathbf{b}) + \sigma^2_n \mathbf{I}_N\right) \mathbf{v} \nonumber \\
\mathbf{q} & = \mathbf{v}^H\mathbf{\Phi}(\mathbf{b})\mathbf{U}\mathbf{D}^{\frac{1}{2}}
\end{align}
It is then easy to show that:
\begin{align}\label{MSE_2}
\mathcal{E}_\mathbf{r}\left(\mathbf{\Lambda}^{(k+1)},\mathbf{\Phi}(\mathbf{b})\right) =\left \| \mathbf{\Lambda}^{(k+1)} \right \|^2 Q \!-\!2\Re \sum\limits_{j=1}^{r}\left[ {\Lambda}^{(k+1)}_j  q_j\right] + r.
\end{align}
The proposed approach to RIS optimization utilizes the Neumann series approximation to linearize the matrix inverses for small variations of the solution, similar to traditional Z-matrix-based approaches (see, e.g., \cite{DR2}). However, as demonstrated in \cite{abrardo_S}, working with the phase of the reflection coefficients rather than directly with the tunable reactances $\mathbf{b}$ leads to better performance in terms of algorithm convergence. For further clarification, when neglecting the parasitic resistance $r_0$, the reflection coefficient $\Gamma_{m}$ of the $m$-th port of the RIS becomes a pure phase shift, i.e., $\Gamma_{m} = e^{j\phi_{m}}$, and we can express it as:
\begin{align}\label{RCoef}
      \Gamma_m = & \frac{jb_m-Z_0}{jb_m+Z_0} \nonumber \rightarrow b_m = \frac{Z_0}{j}\frac{1+e^{j\phi_{m}}}{1-e^{j\phi_{m}}} \nonumber \\
      \frac{d b_m}{d \phi_m} & = 2Z_0\frac{e^{j\phi_{m}}}{\left(e^{j\phi_{m}}-1\right)^2} = \\
      & 2Z_0 \frac{jb_m-Z_0}{\left(jb_m+Z_0\right)\left(\frac{jb_m-Z_0}{jb_m+Z_0}-1\right)^2} = 
      -\frac{b_m^2+Z_0^2}{2 Z_0} \nonumber
\end{align}    
Hence, denote by $\mathbf{B}^{(k)} = \text{diag}\left(\mathbf{b}^{(k)}\right)$ and $\boldsymbol{\phi}^{(k)}$ the tunable reactances matrices and the corresponding reflection coefficient phase vector at iteration $k$ of the iterative algorithm. Considering small variations of $\boldsymbol{\phi}^{(k)}$ for evaluating the solution at next step, i.e., $\boldsymbol{\phi}^{(k+1)} = \boldsymbol{\phi}^{(k)} + \boldsymbol{\delta}$, with $\boldsymbol{\delta} \in \mathbb{C}^{1 \times M}$ and $\delta_m << 1$, we have:
\begin{align}\label{Bvar}
\mathbf{B}^{(k+1)} \approx \mathbf{B}^{(k)} +\mathbf{F}^{(k)}\text{diag}\left(\boldsymbol{\delta}\right)
\end{align}    
where $\mathbf{F}^{(k)} \in \mathbb{C}^{M \times M}$ is a diagonal matrix with entries $\mathbf{F}^{(k)}_{m,m} = -\frac{\left(b_m^{(k)}\right)^2+Z_0^2}{2Z_0}$. 

Since $\boldsymbol{\Phi} = \mathbf{S} \boldsymbol{\Delta}(\mathbf{b})$, from \eqref{eq:TFZ} and \eqref{notations} we have:
\begin{align}\label{notations2}
Q & = 4 Y_0^2 \mathbf{v}^H \left\{\mathbf{S}  (\mathbf{Z}_{SS}+r_0+j\mathbf{B})^{-1} \left(\mathbf{R}_\mathbf{x} +\sum\limits_{i=2}^{N_u} R_{\mathbf{w}_i}\right) \right.\\ 
& \left. \times \left[\mathbf{S}  (\mathbf{Z}_{SS}+r_0+j\mathbf{B})^{-1}\right]^H+ \sigma^2_n \mathbf{I}_N\right\} \mathbf{v} \nonumber \\
\mathbf{q} & = -2Y_0 \mathbf{v}^H\mathbf{S}  (\mathbf{Z}_{SS}+r_0+j\mathbf{B})^{-1}\mathbf{U}\mathbf{D}^{\frac{1}{2}}. \nonumber
\end{align}
From \eqref{Bvar} we have:
\begin{align}\label{refeq1}
& \mathbf{S}  (\mathbf{Z}_{SS}+r_0+j\mathbf{B}^{(k+1})^{-1} \\
& = \mathbf{S}  (\mathbf{Z}_{SS}+r_0+j\mathbf{B}^{(k)} +j\mathbf{F}^{(k)}\text{diag}\left(\boldsymbol{\delta}\right))^{-1}, \nonumber
\end{align}
We then introduce the following terms:
\begin{align}\label{notations3}
 \mathbf{A}^{(k)} & = \left(\mathbf{Z}_{SS}+r_0+j\mathbf{B}^{(k)}\right)^{-1} \\
 \mathbf{G}^{(k)} & = j \mathbf{S}\mathbf{A}^{(k)}  \mathbf{F}^{(k)}. \nonumber
\end{align}
Invoking the Neumann series approximation for matrix inversion, we have:
\begin{align}\label{refeq2}
&\mathbf{S}  (\mathbf{Z}_{SS}+r_0+j\mathbf{B}^{(k+1)})^{-1} \approx \mathbf{S}\mathbf{A}^{(k)} - \mathbf{G}^{(k)}\text{diag}\left(\boldsymbol{\delta}\right)\mathbf{A}^{(k)}
\end{align}
where the approximation holds when:
\begin{align}\label{VNcond}
\left \| \mathbf{F}^{(k)}\text{diag}\left(\boldsymbol{\delta}\right) \left(\mathbf{Z}_{SS}+r_0+j\mathbf{B}^{(k)}\right)^{-1} \right \| \le \epsilon
\end{align}
with $\epsilon << 1$. Denoting by $\mathbf{P} = \mathbf{F}^{(k)} \mathbf{A}^{(k)}$, condition \eqref{VNcond} can be expressed as:
\begin{align}\label{VNcond2}
\sum\limits_{m = 1}^M \delta_m^2 \theta_m^2 < \epsilon^2
\end{align}
where $\theta_m$ represents the entries of the main diagonal of $\mathbf{P}\mathbf{P}^H$.
Then:
\begin{align}\label{notations4}
Q & \approx 4 Y_0^2 \mathbf{v}^H \left\{(\mathbf{S}\mathbf{A}^{(k)} -\mathbf{G}^{(k)}\text{diag}\left(\boldsymbol{\delta}\right)\mathbf{A}^{(k)}) \left(\mathbf{R}_\mathbf{x} +\sum\limits_{i=2}^{N_u} R_{\mathbf{w}_i}\right) \right.\\ 
& \left. \times \left[\mathbf{S}\mathbf{A}^{(k)} -\mathbf{G}^{(k)}\text{diag}\left(\boldsymbol{\delta}\right)\mathbf{A}^{(k)}\right]^H+ \sigma^2_n \mathbf{I}_N\right\} \mathbf{v} \nonumber \\
\mathbf{q} & \approx -2Y_0 \mathbf{v}^H\left(\mathbf{S}\mathbf{A}^{(k)} -\mathbf{G}^{(k)}\text{diag}\left(\boldsymbol{\delta}\right)\mathbf{A}^{(k)}\right)\mathbf{U}\mathbf{D}^{\frac{1}{2}} \nonumber
\end{align}
Introduce now the terms:
\begin{align}\label{notations4}
\mathbf{R} & = \mathbf{R}_\mathbf{x} +\sum\limits_{i=2}^{N_u} R_{\mathbf{w}_i} \nonumber \\
C_1 & = 4 Y_0^2  \mathbf{v}^H\mathbf{S}\mathbf{A}^{(k)} \mathbf{R}\left[\mathbf{A}^{(k)}\right]^H \mathbf{S}^H \mathbf{v} \nonumber \\
C_2 & = 4 \sigma^2_n Y_0^2  \mathbf{v}^H\mathbf{S}\mathbf{A}^{(k)} \mathbf{R} \mathbf{v} \nonumber \\
\mathbf{d} & = -2 Y_0 \mathbf{v}^H  \mathbf{S}\mathbf{A}^{(k)} \mathbf{U}\mathbf{D}^{\frac{1}{2}}
\end{align}
which allows us to reformulate the expression of $Q$ and $\mathbf{q}$ as:
\begin{align}\label{notations5}
Q & \approx  C_1 + C_2 \nonumber \\
& - 2 \Re \left\{4 Y_0^2 \mathbf{v}^H \mathbf{G}^{(k)}\text{diag}\left(\boldsymbol{\delta}\right)\mathbf{A}^{(k)} \mathbf{R} \left[\mathbf{A}^{(k)}\right]^H \mathbf{S}^H \mathbf{v}\right\} \\
& + 4 Y_0^2 \mathbf{v}^H \mathbf{G}^{(k)}\text{diag}\left(\boldsymbol{\delta}\right)\mathbf{A}^{(k)} \mathbf{R}  \left[\mathbf{A}^{(k)}\right]^H \text{diag}\left(\boldsymbol{\delta}\right)  \left[\mathbf{G}^{(k)}\right]^H \mathbf{v} \nonumber \\
\mathbf{q} & \approx \mathbf{d} + 2Y_0 \mathbf{v}^H \mathbf{G}^{(k)}\text{diag}\left(\boldsymbol{\delta}\right)\mathbf{A}^{(k)} \mathbf{U}\mathbf{D}^{\frac{1}{2}} \nonumber
\end{align}
It is now convenient to introduce the terms:
\begin{align}\label{notations6}
&\mathbf{R}_\mathbf{A} = \mathbf{A}^{(k)} \mathbf{R} \left[\mathbf{A}^{(k)}\right]^H \in \mathbb{C}^{M \times M}\nonumber \\
&\mathbf{f}_1 = 4 Y_0^2  \mathbf{v}^H \mathbf{G}^{(k)} \in \mathbb{C}^{1 \times M} \nonumber \\
&\mathbf{f}_2  = \mathbf{R}_\mathbf{A} \mathbf{S}^H \mathbf{v}  \in \mathbb{C}^{M \times 1} \nonumber \\
&\mathbf{f}_3  = \left[\mathbf{G}^{(k)}\right]^H \mathbf{v} \in \mathbb{C}^{M \times 1} \nonumber\\
&\mathbf{f}_4  = 2 Y_0 \mathbf{v}^H\mathbf{G}^{(k)} \in \mathbb{C}^{1 \times M} \nonumber \\ 
&\mathbf{f}_5  = \mathbf{A}^{(k)} \mathbf{U}\mathbf{D}^{\frac{1}{2}} \mathbf{\Lambda}^{(k+1)} \in \mathbb{C}^{M \times 1} \\
&\mathbf{F}_1  = \text{diag}(\mathbf{f}_1) \in \mathbb{C}^{M \times M} \nonumber \\
&\mathbf{F}_3  = \text{diag}(\mathbf{f}_3) \in \mathbb{C}^{M \times M} \nonumber \\
&\mathbf{f}_{1,2}  = \mathbf{f}^T_1 \odot \mathbf{f}_2 \in \mathbb{C}^{M \times 1} \nonumber \\
&\mathbf{f}_{4,5}  = \mathbf{f}^T_4 \odot \mathbf{f}_5 \in \mathbb{C}^{M \times 1} \nonumber 
\end{align}
Accordingly, by considering only the terms in \eqref{MSE_2} that depend on the optimization variable $\boldsymbol{\delta}$, we can approximate the MSE \eqref{MSE_2} as:
\begin{align}\label{MSE_3}
\hat{\mathcal{E}}_\mathbf{r}\left(\mathbf{\Lambda}^{(k+1)},\mathbf{\Phi}(\boldsymbol{\delta})\right) & = \left \| \mathbf{\Lambda}^{(k+1)} \right \|^2 \boldsymbol{\delta} \mathbf{F}_1 \mathbf{R} \mathbf{F}_3   \boldsymbol{\delta}^T \\
& - 2 \boldsymbol{\delta} \left[ \left \|\mathbf{\Lambda}^{(k+1)} \right \|^2 \Re\left( \mathbf{f}_{1,2} \right) + \Re\left( \mathbf{f}_{4,5} \right)\right]  \nonumber
\end{align}
We are now in a position to define the problem for finding the optimal $\boldsymbol{\delta}$ as:
\begin{align}\label{Problem3}
\boldsymbol{\delta}^* = & \text{ arg}\min \limits_{\boldsymbol{\delta}} \hat{\mathcal{E}}_\mathbf{r}\left(\mathbf{\Lambda}^{(k+1)},\mathbf{\Phi}(\boldsymbol{\delta})\right)  \\
&\quad  \textrm{s.t.}
\  \sum\limits_{m = 1}^M \delta_m^2 \theta_m^2 < \epsilon^2 \nonumber 
\end{align}

The problem in \eqref{Problem3} is now convex and can be easily solved in the Lagrangian domain. First, from \eqref{MSE_3} we derive the gradient of the Lagrangian:
\begin{align}\label{gradient11}
& \nabla_{\boldsymbol{\delta}}  \left[\hat{\mathcal{E}}_\mathbf{r}\left(\mathbf{\Lambda}^{(k+1)},\mathbf{\Phi}(\boldsymbol{\delta})\right) + \mu \left(\sum\limits_{m = 1}^M \delta_m^2 \theta_m^2-\epsilon^2\right)\right] \nonumber \\
& = 2\boldsymbol{\delta}  \left \| \mathbf{\Lambda}^{(k+1)} \right \|^2  \Re\left({\mathbf{F}_1 \mathbf{R} \mathbf{F}_3}\right) \\
& - 2  \left \|\mathbf{\Lambda}^{(k+1)} \right \|^2 \Re\left( \mathbf{f}_{1,2} \right) - \Re\left( \mathbf{f}_{4,5} \right)  + 2 \boldsymbol{\delta} \boldsymbol{\Psi} \nonumber
\end{align}
where $\mu > 0$ is a Lagrange multiplier and $\boldsymbol{\Psi} \in \mathbb{C}^{M \times 1}$ is a vector with entries $\theta_m^2$. Accordingly, denoting by $\mathbf{M} = \left \| \mathbf{\Lambda}^{(k+1)} \right \|^2 \Re\left({\mathbf{F}_1 \mathbf{R} \mathbf{F}_3}\right)$ and $\mathbf{v} = \left \|\mathbf{\Lambda}^{(k+1)} \right \|^2 \Re\left( \mathbf{f}_{1,2} \right) + \Re\left( \mathbf{f}_{4,5} \right)$, we finally obtain
\begin{align} \label{opt_2.311}
\boldsymbol{\delta}^* = \left[\mathbf{M}+ \mu \times \text{diag}\left(\boldsymbol{\Psi}\right)\right]^{-1}\mathbf{v}
\end{align} 
where $\mu$ is set to satisfy the constraint in \eqref{Problem3}. Finally, we can evaluate $\mathbf{B}^{(k+1)} \approx \mathbf{B}^{(k)} +\mathbf{F}^{(k)}\text{diag}\left(\boldsymbol{\delta}^*\right)$ and go to the next iteration of the AO algorithm.

\subsubsection{Solution for the CT model} In the CT case, it is necessary to use \eqref{eq:TFZ_CT2} instead of \eqref{eq:TFZ}. This firstly entails using a diagonal $Z_{SS}$ with entries $Z_0$, which is trivial and does not lead to differences regarding the solution of problem \eqref{opt_2_1}. Secondly, it is necessary to subtract $\boldsymbol{\Delta}_S$ in \eqref{refeq2}, resulting in:
\begin{align}\label{refeq2bis}
& \mathbf{S}(\mathbf{Z}_{SS}+r_0+j\mathbf{B}^{(k+1)})^{-1} \\
& \approx \mathbf{S}\mathbf{A}^{(k)} - \mathbf{G}^{(k)}\text{diag}\left(\boldsymbol{\delta}\right)\mathbf{A}^{(k)} -\boldsymbol{\Delta}_S \nonumber.
\end{align}
From this point onwards, the optimization can proceed straightforwardly in the same manner as described above for the MP case.

\section{Numerical Results}
\label{sec:results}
We present some numerical results to assess the effectiveness of the proposed optimization algorithms. Considering the scenario depicted in Fig. \ref{Fig1}, we assume the carrier frequency is $f_0 = 30$ GHz, the RIS is centered at position $P_R = (0,0,3)$ m, and the BS is positioned at $P_{BS} = (-7,7,2)$ m, corresponding to an angle of $-\pi/4$ with respect to the RIS. The $N_u$ UEs are situated at positions $P_{i} = (X_i,Y_i,0)$ m, indicating that they are located on the opposite side of the room from the BS. Specifically, $P_{0}$ is the position of the user of interest and $P_{i}$, $i > 0$ are the positions of the interfering users.

Following the RIS structure proposed in \cite{abrardo_S}, the scatterers of the RIS consist of identical metallic dipoles with a radius of $\lambda/500$ and a length of $L = 0.46 \lambda$.
The dipoles are configured as a uniform planar array (UPA) with $M_H$ elements along the $x$ axis and $M_V$ elements along the $z$ axis. The spacing between the elements is set to $d_x = \lambda/Q$ in the $x$ direction, where $Q$ is an integer determining the dipole density. Conversely, the spacing between elements in the $z$ direction is $d_z = 3/4 \lambda$. We then choose $M_H = 16 \lambda / d_x $ and $M_V = 4$, resulting in an RIS forming a rectangle with an area of $50 \lambda^2$. Regarding the BS, it is equipped with $N = 16$ antennas arranged in two linear arrays along the $z$ axis, with each array containing $8$ antennas along the $x$ axis. The antennas at the BS are separated by $\lambda/2$ in the x-axis and $3\lambda/4$ in the z-axis. Similar to the RIS, these antennas are metallic dipoles with a radius of $\lambda/500$ and a length of $L = 0.46 \lambda$. Regarding the noise, the variance $\sigma^2$ is calculated to achieve a signal-to-noise ratio of -20 dB in the case where the RIS is configured randomly and for a transmitted power $P_{tx} = 20$ dBm. Finally, the UEs are equipped with single dipole antennas having the same characteristics. In the following two subsections, we describe the models used for RIS optimization and for generating simulation results.

\subsection{Model considered for RIS optimization}

Given the potential presence of both Line of Sight (LOS) and non-Line of Sight (NLOS) components in practical channels, we characterize the RIS-UE channel using a generalized spatially correlated Rician fading model \cite{1499046}. This approach is widely used in the literature to account for the presence of multipath fading in RIS-aided communications, as shown, for example, in \cite{9514409} for a multiport network model using S-parameters. 

\begin{figure}[t]
\setlength{\abovecaptionskip}{-0.1cm}
\setlength{\belowcaptionskip}{-0.3cm}   \begin{center}
\includegraphics[scale=0.28]{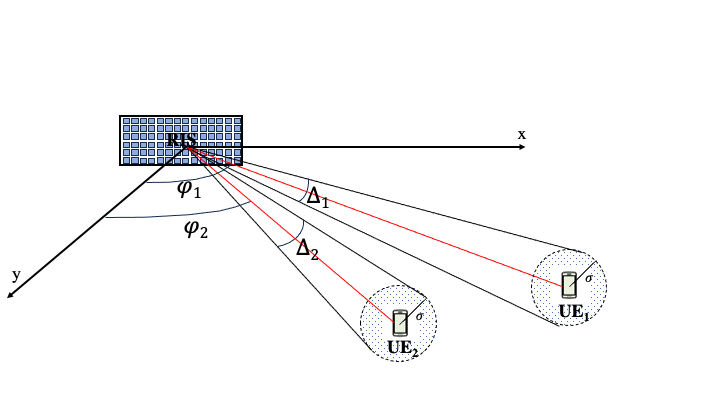}
\end{center}
\captionsetup{font=small}
\caption{Angle spread for correlation evaluation}
\label{Fig2}
\end{figure}

As for the LOS, it is assumed that the prior positions of the nodes are known with circular uncertainty, meaning the nodes are uniformly distrubuted within a circle of radius $\sigma$ around the actual position, corresponding to the localization error. The correlation matrices of the LOS components are then computed using a classical model for correlation calculation based on the angular spread with which the signal can arrive given the known localization error. For simplicity, consider a planar geometry where the arrival angle depends only on the azimuth $\varphi$, and assume knowledge of the range $\Delta_i$ of possible arrival angles for the signal from $i$-th UE, given its uncertainty region (see the illustrative example in Fig. \ref{Fig2}). Denoting by $x_m$ the $x$ location of the $m$-th dipole of the RIS, with $m = 1,\ldots,M$, when a plane-wave impinges on the RIS from the azimuth angle $\varphi$, the array response vector can be written as \cite{Demir2022Channel} 
\begin{align}\label{array_vector_a}
\mathbf{a}(\varphi)=\left[e^{j\frac{2\pi}{\lambda}x_1 \sin \varphi },\ldots,e^{j\frac{2\pi}{\lambda} x_M \sin \varphi}\right]^T,
\end{align}
where $\lambda$ is the wavelength.
Hence, the channel correlation matrices $\mathbf{R}_i$ are evaluated as:
\begin{align}\label{Rh}
\mathbf{R}_i=\beta_i\!\int^{\varphi_i+\Delta_i /2}_{\varphi_i-\Delta_i/2} f(\varphi_i,\Delta_i) \mathbf{a}(\varphi)\mathbf{a}^{\text{H}}(\varphi)d\varphi
\end{align}
where $f(\varphi_i,\Delta_i) = \frac{4(\varphi-\varphi_i)}{\Delta_i}$ is the spatial scattering distribution function which derives from the uniform distribution assumption within the circle, $\beta_i$ is the channel gain which is evaluated assuming a LOS propagation model. Note that this model is equivalent to the spatial correlation matrix model proposed in \cite{Bjornson2021Rayleigh,Demir2022Channel}. However, in this case this model is considered only for CSI-free RIS optimization purposes and does not reflect the model which is generated in the simulations, as explained below. The same model is assumed for the NLOS component. In particular, considering a Ricean factor $K_{1}$ and an angular spread $\Delta_{m}$ and the correlation matrix of the NLOS part is:
\begin{align}\label{Rhmp}
\mathbf{W}_{i}=\frac{\beta_i}{K_1}\!\int^{\varphi_i+\Delta_{m}/2}_{\varphi_i-\Delta_{m}/2}  f(\varphi_i,\Delta_m)  \mathbf{a}(\varphi)\mathbf{a}^{\text{H}}(\varphi)d\varphi
\end{align}
Hence, we evaluate $\mathbf{R}_\mathbf{x} = \mathbf{R}_1+\mathbf{W}_{1}$ and $\mathbf{R}_{\mathbf{w}_i} = \mathbf{R}_i + \mathbf{W}_{i}$, with $i > 1$. 
 
\subsection{Model considered for Simulations} 

Given the estimated positions of the nodes, the true positions are generated according to the uncertainty model, which selects a random point within the circle of the uncertainty region. Meanwhile, the BS is positioned at a fixed location. Subsequently, the Line of Sight (LOS) components of the matrices $\mathbf{S}$ and the vectors $\mathbf{t}_i$, along with the matrix $\mathbf{Z}_{SS}$, are derived from the multiport network model introduced in \cite{DR1}, utilizing the nodes' positions. This model's accuracy has been confirmed in \cite{abrardo_S} through comparison with a full-wave numerical simulator.

To streamline the simulations, we assume that the antennas at the BS are isolated dipoles, neglecting mutual coupling effects at the BS. 
Subsequently, following the Ricean channel model outlined earlier, a Gaussian NLOS component is incorporated. Specifically, if $\mathbf{S}$ and $\mathbf{t}_i$ represent the terms computed by the model \cite{DR1}, the actual $\mathbf{S}^{(e)}$ and $\mathbf{t}_i^{(e)}$ are determined by introducing multipath effects in the form of additive Gaussian noise with a covariance matrix calculated as in \eqref{Rhmp}. Notably, we generate the NLOS for the $i$-th RIS-UE channels, taking into account the angle spreading $\Delta_{m}$ and the Ricean factor $K_1$. For the BS-RIS channel, we consider the angle spreading $\Delta^{(BS)}_{m}$ and a Ricean factor $K_0$. It's worth mentioning that although we include the NLOS component in the BS-RIS channel in the simulations, this effect is not accounted for in the analysis, assuming that $K_0$ can be very high. The direct path BS-UE is not modeled, assuming it is entirely blocked. 

The transfer function used to compute the achievable rate is calculated as in \eqref{eq1} using $\mathbf{S}^{(e)}$ and $\mathbf{t}_i^{(e)}$, and the rate is determined as in \eqref{ERG_Rate}, i.e., with exact channel knowledge. This corresponds to the assumption that given the RIS optimization obtained in an offline phase, during actual transmission, the RIS becomes an integral part of the environment, and communication follows standard end-to-end channel estimation strategies, which can be conducted with pilot sequences and receiver CSI estimation. Thus, the computed rate assumes ideal channel estimation in this phase or negligible errors affecting it.

\subsection{Results} 

In the following figures, we present the results obtained from simulations to validate the CSI-free optimization approach proposed in Section \ref{RIS_opt}. In this regard, the combining vector $\mathbf{v}$ is a beamforming vector. Specifically, denoting $\varphi_{BS}$ as the arrival angle of the signal at the BS from the RIS, $\mathbf{v}$ is calculated as $\mathbf{v} = \mathbf{a}(\varphi_{BS})$, where $\mathbf{a}$ is defined in \eqref{array_vector_a}. Therefore, the results are obtained for two different schemes, namely OPT-NoCSI and CT-NoCSI which correspond to the solution of \eqref{opt_2_1} for the MP, and CT models, respectively. These approaches will be compared with OPT-CSI, which corresponds to the OPT-NoCSI scheme in the case where the channel is exactly known for all involved nodes. This scheme, representing the ultimate upper bound benchmark, can be easily derived from the OPT-NoCSI scheme by calculating $\mathbf{R}_{\mathbf{x}} = \sigma_1^2 \mathbf{t}^{(e)}_1 \left(\mathbf{t}^{(e)}_1\right)^H$ and $\mathbf{R}_{\mathbf{w}_i} = \sigma_i^2 \mathbf{t}^{(e)}_i \left(\mathbf{t}^{(e)}_i\right)^H$. 

\begin{figure}[t]
\begin{center}
\includegraphics[scale=0.4]{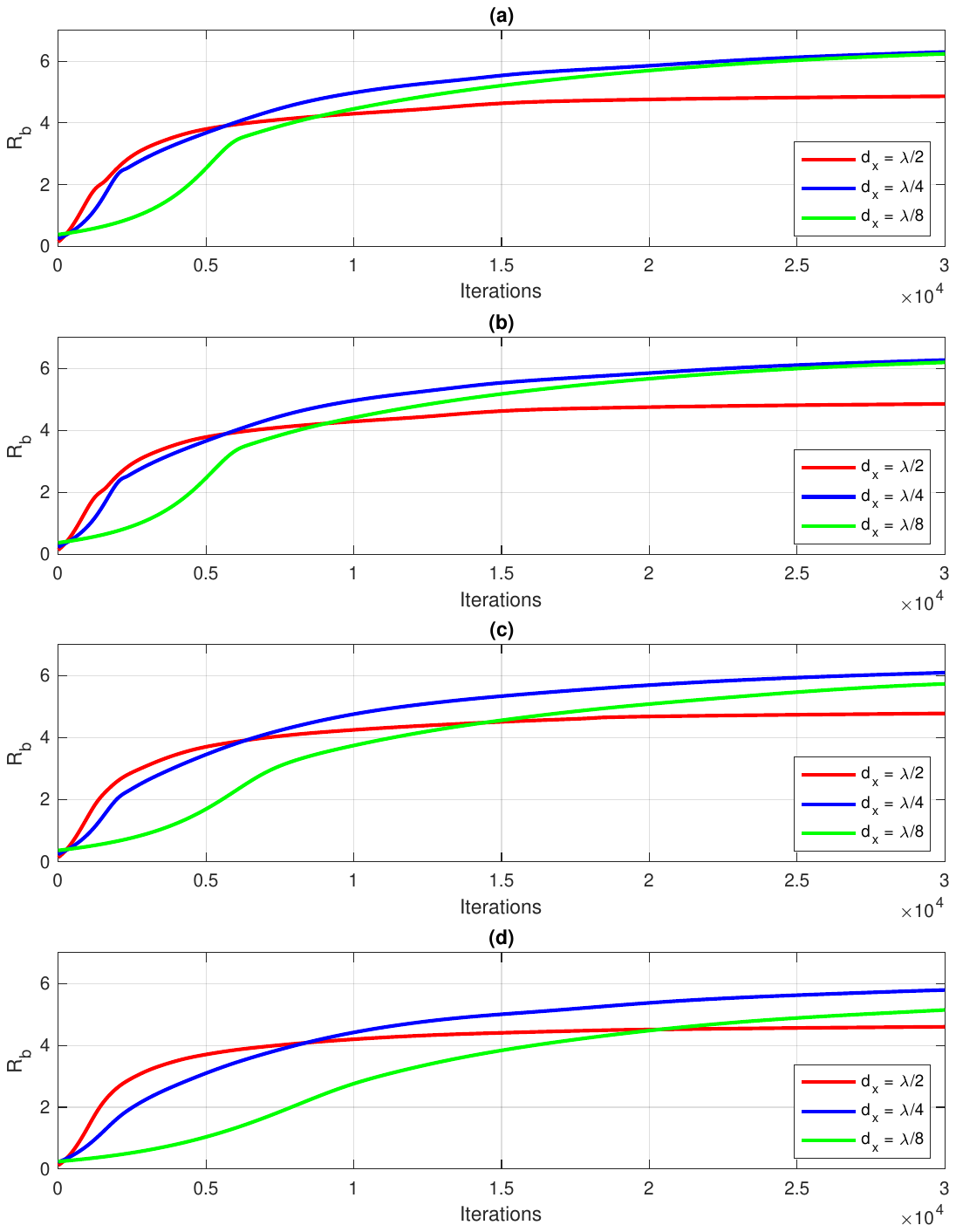}
\end{center}
\captionsetup{skip=-40pt}
\captionsetup{font=small}
\caption{Convergence behavior for the OPT-NoCSI algorithm for the case where $N_u = 3$ and for three values of dipoles' distances: $d_x = \lambda/2$, $d_x = \lambda/4$, and $d_x = \lambda/8$. The node positions are assumed such that the two interfering nodes are located at the same distance of 10 meters from the useful node, with angles between the nodes differing by $\pi/8$. The results for four different position uncertainty, namely $\sigma = 0$ m, $\sigma = 0.1$ m, $\sigma = 0.5$ m and $\sigma = 1$ m, are reported in Figs. (a), (b), (c) and (d), respectively.}
\label{Fig3}
\end{figure}

In the initial figures, the convergence behavior for the OPT-NoCSI algorithm is depicted for the case where $N_u = 3$, with positions $P_0 = 10\left(\cos \pi/8,  \sin \pi/8, 0\right)$, $P_1 = 10\left(\cos \pi/4, \sin \pi/4, 0\right)$, and $P_2 = 10\left(\cos 3\pi/8, \sin 3\pi/8, 0\right)$, and for three values of dipoles' distances: $d_x = \lambda/2$, $d_x = \lambda/4$, and $d_x = \lambda/8$. Specifically, the rate value \eqref{ERG_Rate1} expressed in bit/s/Hz is plotted against the iteration index $k$. In other words, the node positions are assumed such that the two interfering nodes are located at the same distance of 10 meters from the useful node, corresponding to the case where the signal-to-interference ratio (SIR) is equal to 1, with angles between the nodes differing by $\pi/8$. The Ricean factors $K_{0}$ and $K_1$ for the BS-RIS and RIS-UEs links are set to $K_0 = 20$ and $K_1 = 10$, respectively. Figure \ref{Fig3} (a), (b), (c) and (d) report the results for four different position uncertainty: $\sigma = 0$ m, $\sigma = 0.1$ m, $\sigma = 0.5$ m and $\sigma = 1$ m, respectively.
It can be observed that the proposed algorithm allows for an increasing rate $R_b$ with iterations, thus enabling convergence to a local optimum in all cases. In this regard, it is noted that when the dipoles are spaced by $\lambda/2$, convergence is faster, which can be explained by the lesser effect of the non-diagonals of the Z matrix. On the other hand, better performance is observed when the dipoles are closer, an effect already highlighted in previous works using multiport network models. It is noted that this improvement is achieved for the same RIS aperture of $50 \lambda^2$. Finally, it is observed that increasing position uncertainty results in only a slightly perceptible performance degradation. For example, in the case of $\lambda/8$, the rate decreases from approximately 6.5 to 6. It should be emphasized that this is not the actual rate measured by the simulations and reported in the subsequent graphs, but only the rate used in the objective function to optimize the RIS in the offline phase.

\begin{figure}[t]
 \begin{center}
\includegraphics[scale=0.48]{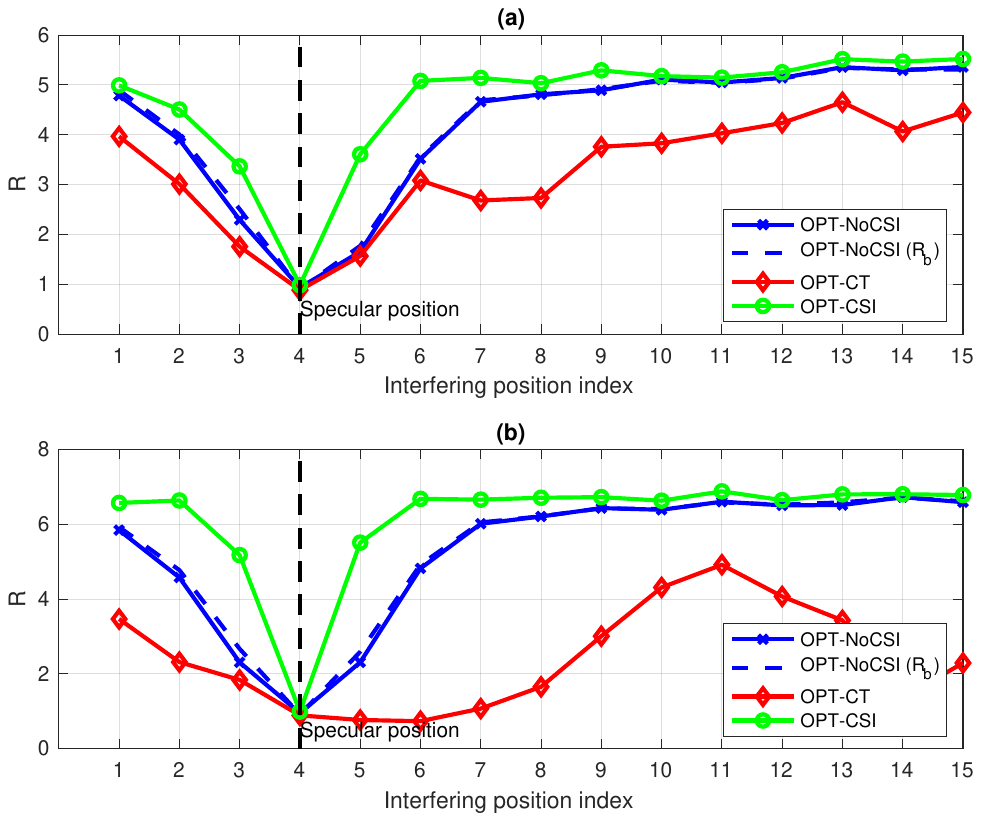}
\end{center}
\captionsetup{skip=-80pt}
\captionsetup{font=small}
\caption{Rate $R$ in \eqref{ERG_Rate} evaluated through simulations considering $\sigma = 0.5$ m in the case of $N_u = 2$, with the intended UE positioned at $P_0 = 10\left(\cos \frac{\pi}{8}, \sin \frac{\pi}{8}, 0\right)$ while the interfering user is positioned at 12 possible different positions, $P_i = 10\left\{\cos [\pi/8 + (i-1)\pi/32], \sin[\pi/8 + (i-1)\pi/32],0\right\}$, with $i = 1,\ldots,12$.  The case (a) refers to $d_x = \frac{\lambda}{2}$, while the case (b) reports the case $d_x = \frac{\lambda}{4}$}
\label{Fig1_rate}
\end{figure}

\begin{figure}[t]
 \begin{center}
\includegraphics[scale=0.48]{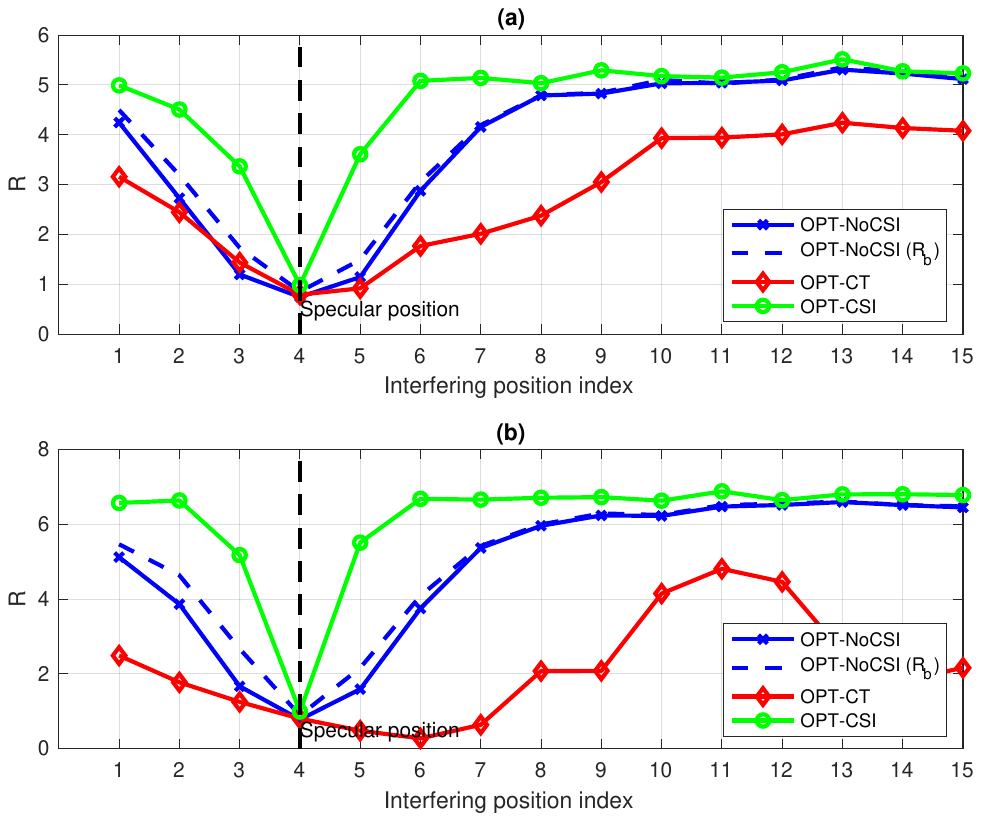}
\end{center}
\captionsetup{skip=-80pt}
\captionsetup{font=small}
\caption{Rate $R$ in \eqref{ERG_Rate} evaluated through simulations considering $\sigma = 1$ m for the same setting of Fig. \ref{Fig1_rate}.}
\label{Fig2_rate}
\end{figure}

\begin{figure}[t]
 \begin{center}
\includegraphics[scale=0.48]{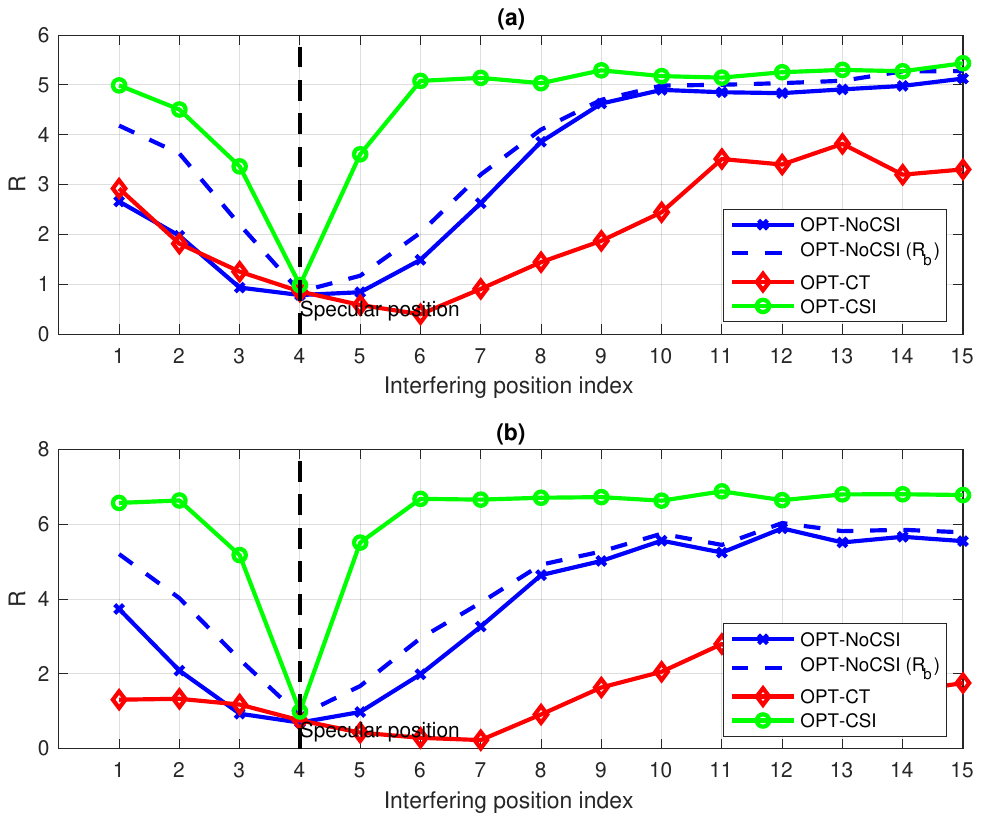}
\end{center}
\captionsetup{skip=-80pt}
\captionsetup{font=small}
\caption{Rate $R$ in \eqref{ERG_Rate} evaluated through simulations considering $\sigma = 2$ m for the same setting of Fig. \ref{Fig1_rate}.}
\label{Fig3_rate}
\end{figure}

In Figs. \ref{Fig1_rate}-\ref{Fig3_rate}, the trend of the rate $R$ in \eqref{ERG_Rate} evaluated through simulations is depicted, considering $\sigma = 0.5$ m in the case of $N_u = 2$ with the intended UE positioned at $P_0 = 10\left(\cos \pi/8, \sin \pi/8, 0\right)$ while the interfering UE is positioned at 12 different possible positions $P_1 = T_i = 10\left[\cos (i \times \pi/32), \sin(i \times \pi/32)\right]$ with $i = 1,2,\ldots,15$. Accordingly, the interfering user is at the same distance from the RIS as the useful user but with different angles. In Figs. \ref{Fig1_rate}-\ref{Fig3_rate}(a), the case $d_x = \lambda/2$ is considered, whereas in \ref{Fig1_rate}-\ref{Fig3_rate}(b) $d_x = \lambda/4$ is reported. Furthermore, the three figures \ref{Fig1_rate}, \ref{Fig2_rate}, and \ref{Fig3_rate} refer to the values of $\sigma = 0.5,1,2$ m, respectively. The curves correspond to the OPT-NoCSI, OPT-CT, and OPT-CSI cases described earlier. It is noted that at position $T_4$, the interfering user is on average in the same position as the intended UE, thus an average rate equal to one is expected since SIR = $1$ and the interference cannot be removed, as indeed happens in all considered cases. It is also noted that at position $T_8$, the interfering user is at an angle $\pi/4$, corresponding to the specular position of the RIS with respect to the BS. In this case, it is evident that the CT model fails, as the RIS is optimized without considering the specular component, which in this case produces strong interference. The dashed curve also reports the value of $R_b$ defined in \eqref{ERG_Rate1}, which corresponds to the objective used by the optimizer. It is noted how this value closely approximates the rate $R$ found through simulations in all cases. Finally, it is noted that the optimizer OPT-NoCSI, in many cases, even for relevant position estimation errors, manages to achieve performance reasonably close to the OPT-CSI case, which requires an exact channel estimation and, in any case, provides much better performance than the OPT-CT case, especially in the presence of consistent mutual coupling (i.e., for $d_x = \lambda/4$), and when the interferer is close to the specular angular position.


\begin{figure}[t]
 \begin{center}
\includegraphics[scale=0.35]{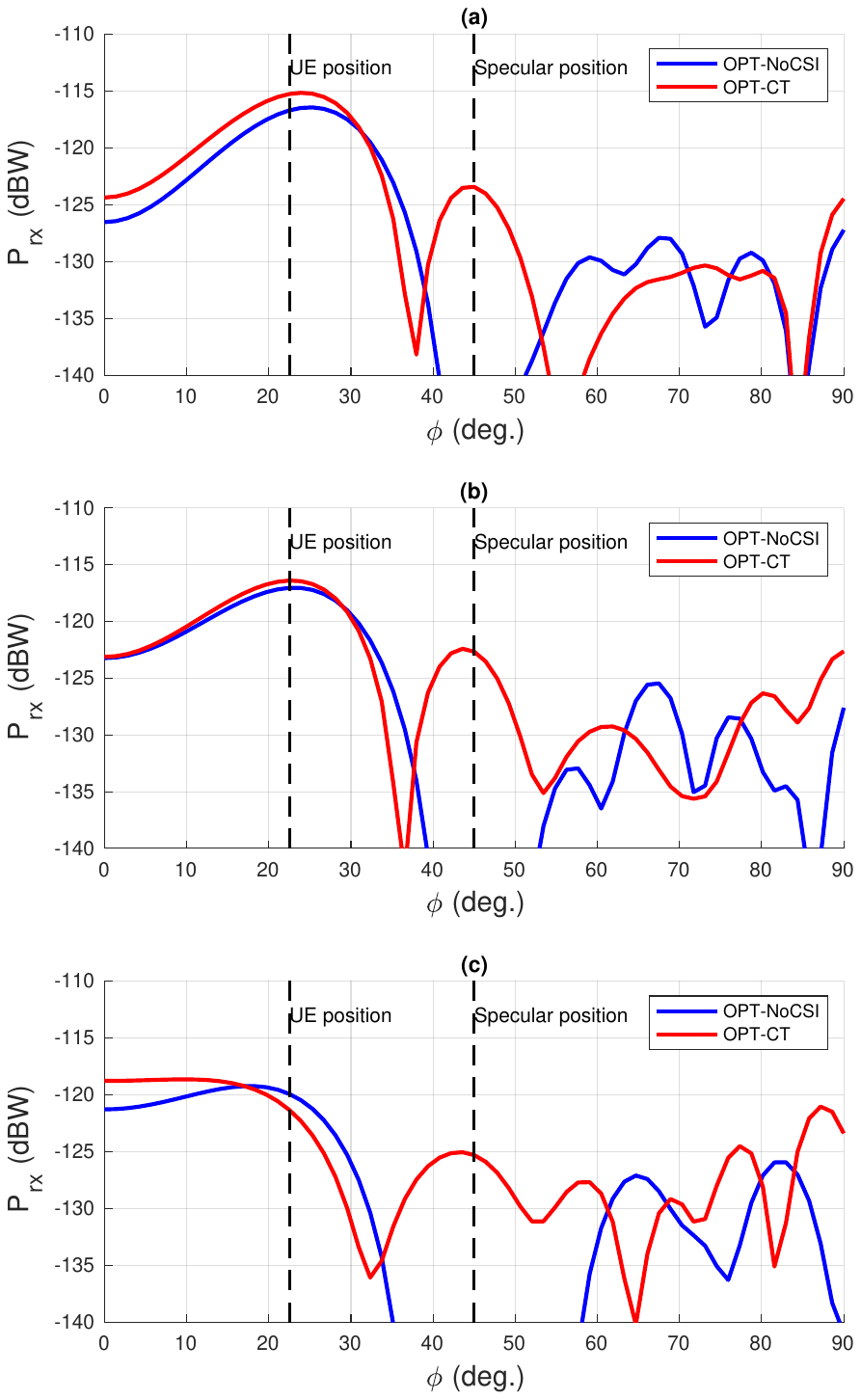}
\end{center}
\captionsetup{skip=-20pt}
\captionsetup{font=small}
\caption{Power received at the BS (in dBW) as a function of the angle $\phi$ of the transmitting node. The RIS characterized by $d_x = \lambda/2$ and it is optimized using the OPT-NoCSI and CT approaches for the case $N_u = 2$, where the reference node at an angle of 22.5 degrees, and the interfering node is at 45 degrees (specular direction). }
\label{Fig1_diag}
\end{figure}

\begin{figure}[t]
 \begin{center}
\includegraphics[scale=0.35]{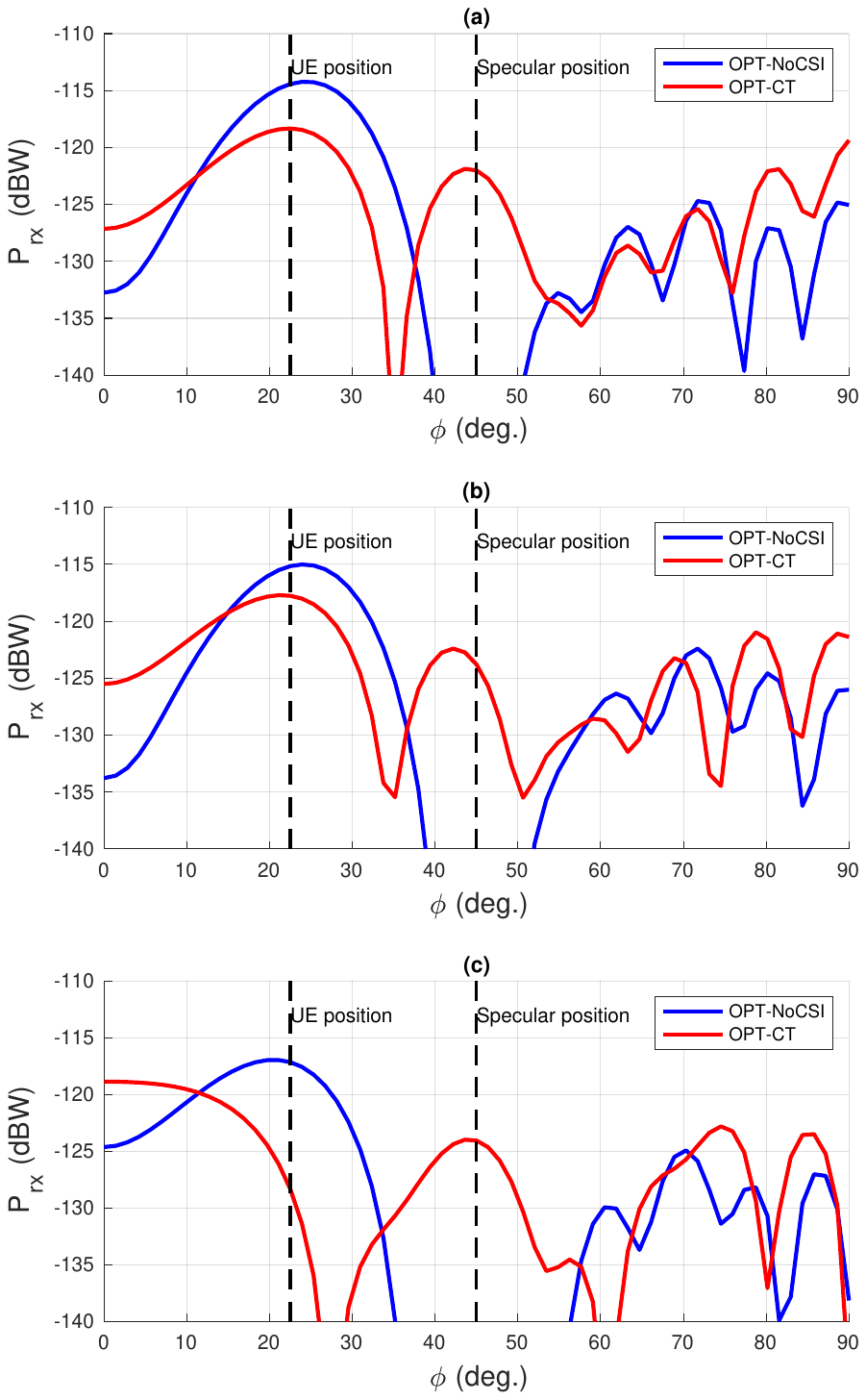}
\end{center}
\captionsetup{skip=-20pt}
\captionsetup{font=small}
\caption{Power received at the BS (in dBW) as a function of the angle $\phi$ of the transmitting node for $d_x = \lambda/4$ and the same setting as Fig. \ref{Fig1_diag}.}
\label{Fig2_diag}
\end{figure}

In the last two figures, namely \ref{Fig1_diag} and \ref{Fig2_diag}, we present the scenario where there is only one transmitting node positioned at a distance of 10 m and at all possible angular positions $\phi$ from 0 to 90 degrees. In this setting, we depict the power received at the BS (in dBW) as a function of the angle $\phi$. The RIS is optimized using the OPT-NoCSI and CT approaches for the case $N_u = 2$, where the central position of the reference node is $P_0 = 10\left(\cos \frac{\pi}{8}, \sin \frac{\pi}{8}, 0\right)$, corresponding to an angle of 22.5 degrees, and that of the interfering node is $P_1 = 10\left(\cos \frac{\pi}{4}, \sin \frac{\pi}{4}, 0\right)$, corresponding to the specular position at 45 degrees. In \ref{Fig1_diag} (a), (b), (c), the cases $d_x = \lambda/2$ and $\sigma = 0.5,1,2$ m are depicted, respectively. In \ref{Fig2_diag} (a), (b), (c), the same cases are illustrated for $d_x = \lambda/4$. The objective of these figures is to showcase the type of beam created by the RIS in the direction of the useful node and its ability to reject interference.

The first notable observation is that in the CT case, the interference is not eliminated. This occurs because in this model, structural scattering is not considered, and thus, the interference originating from the direction $\pi/4$ is significantly underestimated. Conversely, in the OPT-NoCSI case, the RIS places a null in the direction of the specular component. In both cases, the RIS generates a beam in the direction of the useful signal. However, in the CT case when $d_x = \lambda/4$, the beam is not centered due to the mismatch of the model, which does not account for mutual couplings. Both the null towards the interferer and the beam towards the intended UE become wider as the uncertainty regarding the possible position of the interferer increases (observe the difference in the beams when transitioning from case (a) to case (c)). This behavior aligns with expectations, as an increase in position uncertainty necessitates 'illuminating' or 'eliminating' larger portions of space with the RIS.

 

\section{Conclusion}

In this paper, we have addressed the challenge of optimizing Reconfigurable Intelligent Surfaces (RIS) for uplink communication systems in the presence of interfering users without relying on Channel State Information (CSI) estimation. Our proposed CSI-free approach leverages a priori statistical knowledge of the channel to maximize the average achievable rate. To account for practical considerations often overlooked in traditional RIS models, we developed a multiport network model that incorporates factors such as mutual coupling among scattering elements and the presence of structural scattering. Through simulations, we have demonstrated the effectiveness of our approach, showing that it can achieve performance levels close to those obtained with perfect CSI in some cases.


\end{document}